\documentclass[
 amsmath, amssymb, breqn,
 aps, prl, superscriptaddress, twocolumn,
longbibliography]{revtex4-1}

\usepackage{graphicx}
\usepackage{amsmath,amssymb}
\usepackage{textcomp}

\newcommand{\dd}{\ensuremath{\mathrm{d}}}

\newcommand{\diff}[2]{\ensuremath{\frac{\dd {#1}}{\dd {#2}}}}
\newcommand{\pdiff}[2]{\ensuremath{\frac{\partial {#1}}{\partial {#2}}}}

\newcommand{\bv}[1]{\ensuremath{\mathbf{#1}}}

\newcommand{\Fes}{\ensuremath{F_\mathrm{es}}}

\def\3He{$^3$He}
\def\4He{$^4$He}

\begin{document} 

\title{Observation of bistable turbulence in quasi-two-dimensional superflow}

\author{E. Varga}
\author{V. Vadakkumbatt}
\author{A. J. Shook}
\author{P. H. Kim}
\author{J. P. Davis}
\affiliation{Department of Physics, University of Alberta, Edmonton, Alberta T6G 2E1, Canada}

\begin{abstract}
  Turbulent flow restricted to two dimensions can spontaneously develop order on large scales, defying entropy expectations and in sharp contrast with turbulence in three dimensions where nonlinear turbulent processes act to destroy large-scale order. In this work we report the observation of unusual turbulent behavior in steady-state flow of superfluid \4He---a liquid with vanishing viscosity and discrete vorticity---in a nearly two-dimensional channel. Surprisingly, for a range of experimental parameters, turbulence is observed to exist in two bistable states. This bistability can be well explained by the appearance of large-scale regions of flow of opposite vorticity. 
\end{abstract}

\maketitle 

Chaotic motion of flowing fluids---turbulence---is one of the most ubiquitous phenomena occurring in nature and is frequently encountered in everyday life. Typically, the turbulence that one encounters takes place in three dimensions (3D), however, two dimensional (2D) turbulence, while not perfectly realized in nature, is relevant to systems where motion in two dimensions dominates over the third, such as large-scale flows in oceans \cite{Arbic2013}, atmospheres \cite{Pouquet2013}, soap bubbles \cite{Rivera1998} or liquid crystal films \cite{Tsai2004}.  The hallmark feature of turbulence in three dimensions is the transfer of energy from large scales to small scales in both classical \cite{Frisch_book} and quantum \cite{Navon2019} fluids. This ``Kolmogorov cascade'' can be understood as the splitting of large eddies in the flow into progressively smaller ones, until viscous damping dominates and dissipates the kinetic energy of small scales into heat. Turbulence in 3D therefore acts to destroy any large scale ordering and, indeed, homogeneous and isotropic turbulence is an excellent approximation in many cases.

Restricting the flow of a classical fluid to 2D disrupts this homogenizing behavior. Interestingly, the direction of the cascade of energy can be inversed \cite{Kraichnan1980,Boffetta2012} and vorticity can coalesce into large eddies, thus spontaneously generating large-scale order from forcing on smaller scales. If the vorticity of the system is discrete (e.g., the quantized vortices in Bose-Einstein condensates \cite{Gauthier2019,Johnstone2019} or the superfluid phases of \3He and \4He \cite{Sachkou2019}, as opposed to continuous vorticity of classical fluids), one can treat the system as a `gas' of point-like vortices, which can be analyzed using the tools of statistical mechanics. In his pioneering work, Onsager \cite{Onsager1949} showed that such a gas can exist at effectively negative temperatures, which would physically manifest as clusters of like-signed vortices (i.e., configurations with high energy and low entropy), similar to the large-scale eddies in 2D classical fluids.

Sixty years after its prediction, the Onsager vortex gas has recently been observed: first using quantized vortices in Bose-Einstein condensates (BECs) \cite{Gauthier2019,Johnstone2019} and then using a nanometer-thick film of superfluid \4He \cite{Sachkou2019}. These systems, however, contained only a small number of vortices ($N < 50$) and were allowed to decay freely during the experiment. Therefore open questions remain as to the robustness of this phenomenon in macroscopic systems with large number of vortices and in steady-state flows (regimes approached so far only in simulations \cite{Reeves2017,Reeves2013a}). In this work, we study a forced and strongly turbulent oscillatory flow in a \textmu m-thick slab of superfluid \4He with macroscopic (mm-scale) lateral size. Turbulence in this system can exist in two nearly-degenerate bistable states, both different from the laminar (i.e., non-turbulent) state. The transitions between these individual flow states are discontinuous, hysteretic, and a highly unusual ``backward'' transition from a less-turbulent to more-turbulent state upon decrease in velocity is observed. We argue that these observations stem from quasi-2D physics and that both the bistability and backward transition are naturally explained in terms of spontaneous flow polarization, suggesting the presence of large-scale order.

\begin{figure}[b]
      \includegraphics{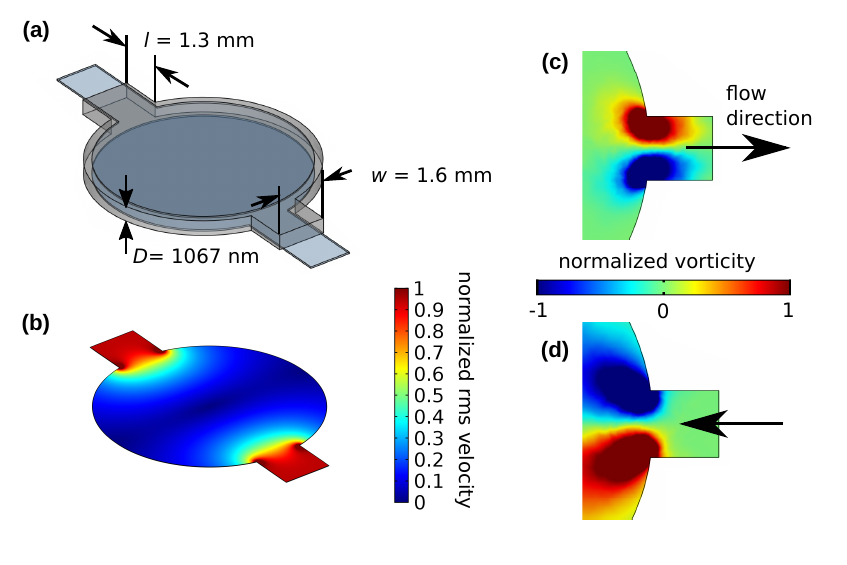}
  \caption{Polarized vorticity in a 2D Helmholtz resonator. 
  (a) Sketch of the device. The central circular basin, which is only used to drive and sense the flow, is connected to the surrounding bath of He-II through two side channels. 
  (b) Simulation of the Helmholtz mechanical mode, showing the normalized root-mean-square velocity concentrated in the channels. 
  (c,d) Vorticity is induced by sharp corners during the forward and reverse stroke of the mechanical mode; arrow indicates the superfluid flow direction. 
  The positive and negative vortices are well separated in space (cf.~the vortex generation terms $g_\pm$ in Eqs.~\ref{eq:dnpdt},\ref{eq:dnmdt}) and the injection of vorticity into the channel is thus strongly polarized (cf.~the vortex polarization generation term $g_s$ in Eq.~\ref{eq:dsdt}).
  }
  \label{fig:flow}
\end{figure}
We study turbulence in superfluid \4He (He-II), which behaves as a mixture of two distinct fluid components \cite{Tilley_book}---an inviscid superfluid, where vorticity is restricted to discrete quantized vortices, and a normal fluid, with continuous vorticity and finite viscosity. He-II has proven to be a valuable testbed \cite{Barenghi2014c,Walmsley2014a,Fonda2019,Baggaley2014b} for the study of turbulence with both continuous and discrete vorticity, as well as the interactions between them. 

Here, oscillatory flow is excited inside a microfluidic Helmholtz resonator \cite{Rojas2015,Souris2017,Shook2020} immersed in He-II, where flow is limited to nearly two dimensions by confinement in the vertical direction  ($D = 1067$ nm, Fig.~\ref{fig:flow}(a)). A uniform confinement, while somewhat larger than the thickness of previously used adsorbed films \cite{Sachkou2019}, avoids dissipative effects stemming from vortex-surface interactions \cite{Forstner2019}. Due to this strong confinement, only the superfluid component of He-II can move (the normal component being viscously clamped \cite{Souris2017}). The resonator is microfabricated from single-crystal quartz and consists of a central circular basin connected to a surrounding bath of He-II through two equal opposing channels of rectangular cross-section (Fig.~\ref{fig:flow}(a)). The capacitive driving and sensing of the flow (see \cite{Rojas2015,Souris2017,Shook2020} and \cite{SM} for more details) allows us to measure the relationship between the driving pressure gradient and the fluid velocity in the channel of the resonator. The confinement used in this study was 1067 nm, but qualitatively similar results were also obtained for 805 nm confinement (see \cite{SM}).

A simulation of the fluid Helmholtz mechanical mode, Fig.~\ref{fig:flow}(b), shows that the flow velocity is essentially confined to the two side channels. As the fluid flows into or out of the channel, the sharp corners at the channel end induce a vorticity in the flow, as seen in Fig.~\ref{fig:flow}(c,d). On the forward stroke (fluid flowing into the channel, Fig.~\ref{fig:flow}(c)) vorticity is injected into the channel in a polarized fashion. On the reverse stroke (Fig.~\ref{fig:flow}(d)) vorticity is ejected and lost into the basin. The two side channels are identical, thus any flow instabilities are likely to occur approximately simultaneously.

\begin{figure}
  \centering
  \includegraphics{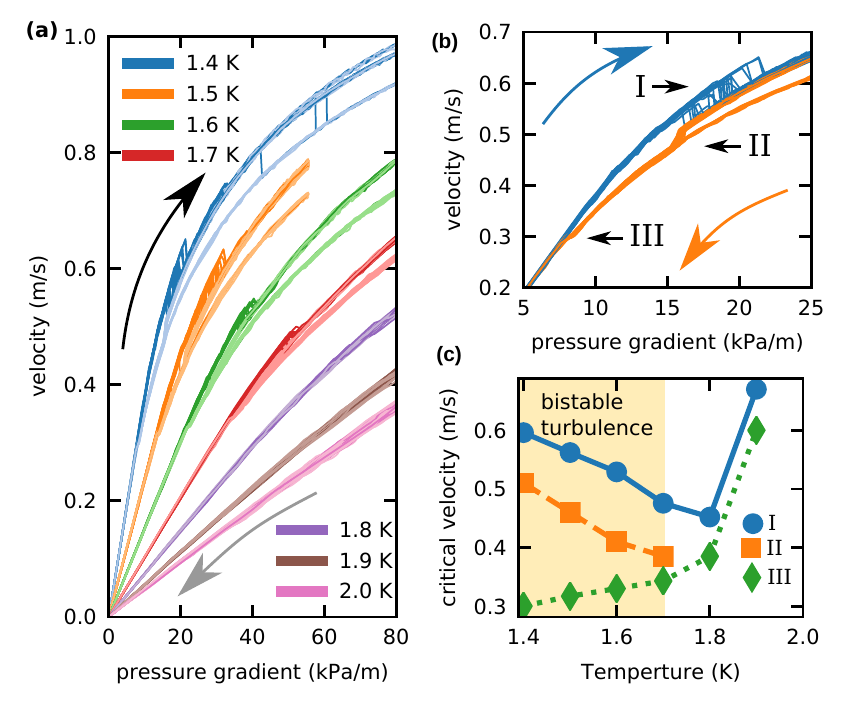}
  \caption{Bistable turbulence and critical velocities. 
  (a) The measured flow velocity as a function of applied pressure gradient for a range of temperatures. Darker curves show increasing pressure gradient, lighter decreasing (as indicated by the curved arrows). 
  (b) Detail of the laminar-to-turbulent transition at 1.4 K. Blue curves correspond to increasing drive, orange to decreasing drive. Three critical velocities with discontinuous jumps are apparent: ``I" -- transition from laminar to turbulent state upon increasing velocity; ``II" -- the unusual backward transition into a more dissipative state upon decrease in velocity; ``III" -- transition from turbulent state back to laminar flow. Above ``I", the flow can randomly transition into the more dissipative state (not shown, see \cite{SM}).
  (c) The temperature dependence of the critical velocities of the three types labeled in panel (b). Onset of the bistable turbulence coincides with the appearance of the critical velocity ``II" of the backward transition.
  }
  \label{fig:experimental}
\end{figure}

 To study the dissipation in this quasi-2D flow we resonantly drive the Helmholtz mechanical mode and continuously increase or decrease the drive amplitude (no significant dependence on ramp rate was observed). Several repeated measurements of peak velocity as a function of peak applied pressure at nominally identical conditions are shown in Figs.~\ref{fig:experimental}(a)-\ref{fig:experimental}(b). For small drives the behavior is linear (i.e., the flow is laminar). With increasing drive, however, the measured velocity falls short of the value expected by extrapolating from the linear regime. That is, above a critical velocity the flow is damped by a drag with nonlinear dependence on velocity. The damping in the linear regime is believed to be dominated by the thermoviscous effect \cite{Souris2017}, whereas the nonlinear damping is predominantly due to presence of quantized vortices \cite{Gao2018a}. The transition to the nonlinear regime is hysteretic and is marked by a discontinuous jump in the velocity-pressure dependence. For temperatures below 1.7 K, the velocity-pressure dependence in the nonlinear regime randomly follows one of two distinct and well-defined curves, i.e., the turbulence is bistable.

The temperature dependence of the observed critical velocities (defined as the mean positions of the discontinuous jumps, see Fig.~\ref{fig:experimental}(b)) is shown in Fig.~\ref{fig:experimental}(c). Here, a new critical velocity---type ``II"---appears below 1.7 K, which coincides with beginning of the bistable regime. In this bistable regime, as the flow velocity decreases the intermediate turbulent state with lower dissipation (i.e., higher velocity at a given drive) destabilises but, rather than becoming laminar again, the flow transitions into the state with stronger turbulence (i.e., lower velocity at a given drive), as can be seen in Fig.~\ref{fig:experimental}(b). This results in a highly unusual ``backward'' transition (critical velocity ``II'' in Fig.~\ref{fig:experimental}(b)) into a state with \emph{higher} dissipation as the flow velocity \emph{decreases}.

The microscopic confinement and large aspect ratio of our flow suggests the use of a 2D theory such as the Onsager vortex gas model \cite{Onsager1949,Kraichnan1980}. However, our system deviates from the Onsager model in several important aspects: it is dissipative, continuously driven, and the confinement is large compared to the thickness of a quantized vortex ($\approx 10^{-10}$ m). Since our measurements have been conducted at relatively high temperatures, mutual friction will strongly attenuate any highly-curved vortex structures. Therefore, for the sake of simplicity of the modelling, we assume that the majority of the vortices in our system can be described by two populations with definite orientations, i.e., the vortices are approximately point-like (see \cite{SM} for more detailed analysis). We note, however, that a population of vortices without definite polarization (i.e., loops attached to a single wall) almost certainly exists in our system. In quasi-2D modelling these can be approximated as point-vortex dipoles. Furthermore, our experiment is sensitive to the total dissipation, which is an integral quantity, and hence we cannot directly determine the presence of, e.g., negative vortex temperatures, for which we would need to know the positions and signs of the vortices \cite{Groszek2018a,Valani2018}. However, the spatial separation of vortices of differing signs explains the observed bistability, hysteresis, and backward transition.

To show this, we construct a model for the number of vortices in the system that captures the essential physics. A similar approach has been adapted for 2D BECs \cite{Groszek2016,Cidrim2016} and 3D counterflow of He-II \cite{Varga2018}. We model the time evolution (on time-scales long comparable to the flow oscillation period) of positively- and negatively-oriented local vortex densities, $n_+$ and $n_-$, respectively as
\begin{equation}
  \label{eq:dnpdt}
  \pdiff{n_+}{t} = an_+ + bn_- - n_+n_-d + g_+,
\end{equation}
\begin{equation}
  \label{eq:dnmdt}
  \pdiff{n_-}{t} = an_- + bn_+  - n_+n_-d + g_-.
\end{equation}
Here, the terms on the right-hand-side correspond to removal of vortices by advection ($a < 0$), creation of new vortices by splitting of seed vortices ($b > 0$), annihilation of a pair of vortices of opposite orientation ($d > 0$), and creation of vortices by large scale shear ($g_\pm > 0$). We note in passing that these equations are similar to the Lotka-Volterra, or predator-prey, equations used to model population dynamics in ecology \cite{Berryman1992}, oscillatory chemical reactions \cite{Lotka1910} or, indeed, the transition to turbulence \cite{Shih2016}. Restricting the model to total vortex density $n = n_+ + n_-$ and polarization $s = (n_+ - n_-)/n$ we have
\begin{equation}
  \label{eq:dndt}
  \pdiff{n}{t} = (a+b)n - \frac{1}{2}dn^2(1 - s^2) + g,
\end{equation}
\begin{equation}
  \label{eq:dsdt}
  \pdiff{s}{t} = -2bs + \frac{1}{2}dns(1 - s^2) + \frac{g_s}{n},
\end{equation}
where $g = g_+ + g_-$ and $g_s = [(1-s)g_+ - (1 + s)g_-]$, which we take as our control parameters and assume their independence of $s$ and $n$ (see \cite{SM} for details). The term $g_s$ represents the polarization of the drive, i.e., the separation of generation of positive and negative vortices on the opposing corners of the channel (see Fig.~\ref{fig:flow}). The vortex densities $n_\pm$ and vortex generation terms $g_\pm$ are local, whereas only the total drag, determined by the total number of vortices $n$, is measured in the experiment. As a first approximation we can replace the density $n$ by its spatial average. The polarization $s$ is anti-symmetric with respect to the device axis (see Fig.~\ref{fig:flow}) and its average vanishes, assuming that the flow remains neutral. Therefore we decompose $s(\bv r)$ into a series of appropriate orthogonal modes $s(\bv r) = \sum_k s_k(\bv r)$. Truncating the expansion after the leading term, we use Eq.~\ref{eq:dsdt} for calculating the evolution of a single mode which captures the large-scale polarization of the vortex distribution.

\begin{figure}[t]
  \centering
  \includegraphics{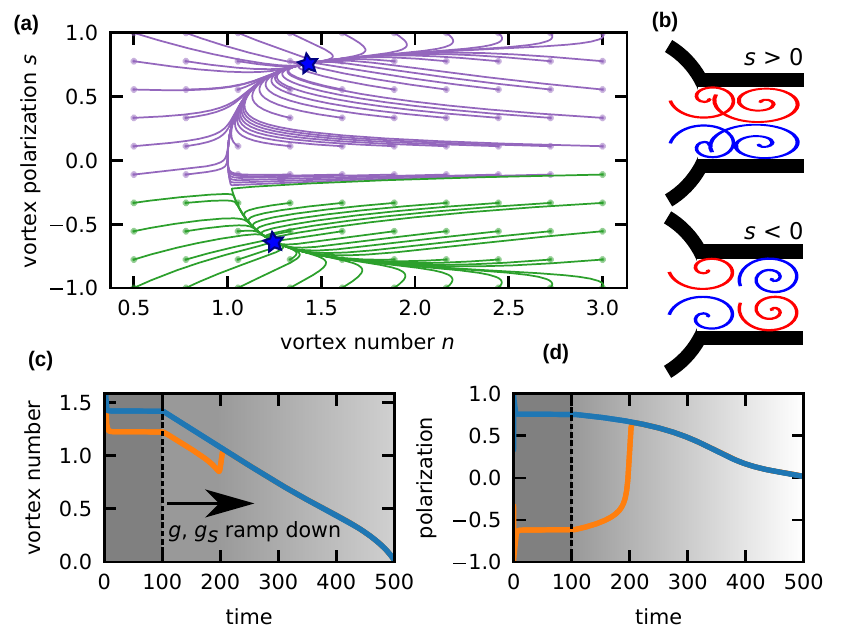}
  \caption{Bistability of the quasi-2D model, Eqs.~\ref{eq:dndt}, \ref{eq:dsdt}, for an example set of parameters $a = -1$, $b = 0.5$, $d = 3$, $g = 2$, and $g_s = 0.1$. 
  (a) Starting from initial conditions indicated by the filled circles, the solution to equations \ref{eq:dndt}, \ref{eq:dsdt} approaches (indicated by the color of the trajectory) one of the stationary points shown as the blue stars.
  (b) Illustration of the two turbulent states. The large-scale polarization of the flow is either aligned ($s > 0$) or anti-aligned ($s<0$) with the polarization of the drive (cf.~Fig.~\ref{fig:flow}(c)).
  (c,d) The backward transition in $n$ (c) and $s$ (d) during linear (in time) ramp-down of the generation parameters $g$ and $g_s$ in the range $100 < t < 500$. The flow is preferentially driven into the $s>0$ state. For sufficiently large number of vortices $n$, however, $s<0$ is also stable (i.e., the flow state can absorb oppositely oriented vortices without collapsing). As the drive---and thus the vortex number---decreases, the $s < 0$ state becomes unstable and the flow switches to the $s>0$ state. The drive and flow are now aligned, thus the vortex number increases.}
  \label{fig:model}
\end{figure}

The dynamical system of Eqs.~\ref{eq:dndt}, \ref{eq:dsdt} indeed has two stationary solutions for certain choices of parameters that differ in both $s$ and $n$, as shown in Fig.~\ref{fig:model}(a). The essential reason for the existence of two distinct steady states of vortex number $n$ is that $g_s \neq 0$ (i.e., the drive is polarized, see Fig.~\ref{fig:flow}(c) and Fig.~\ref{fig:model}(b)), which lifts the degeneracy of $s > 0$ and $s < 0$ stationary solutions, when they exist. 

By starting the evolution of the system from either of the stationary solutions and switching off the generation terms $g, g_s$ linearly in time, we model the velocity ramp-down experiment. The result, shown in Figs.~\ref{fig:model}(c)-~\ref{fig:model}(d), reproduces the unusual backward transition observed experimentally. The transition occurs due to destabilisation of the $s<0$ polarization state, which is stabilised only at sufficiently high vortex densities $n$ (i.e., the flow state is robust enough to withstand the oppositely polarized drive). As the drive, and the overall vortex number, decreases, this state destabilizes and transitions into the $s > 0$ state. When the flow and drive polarizations are aligned, fewer vortices are annihilated and thus the vortex density $n$ temporarily increases.

\begin{figure}
    \centering
    \includegraphics{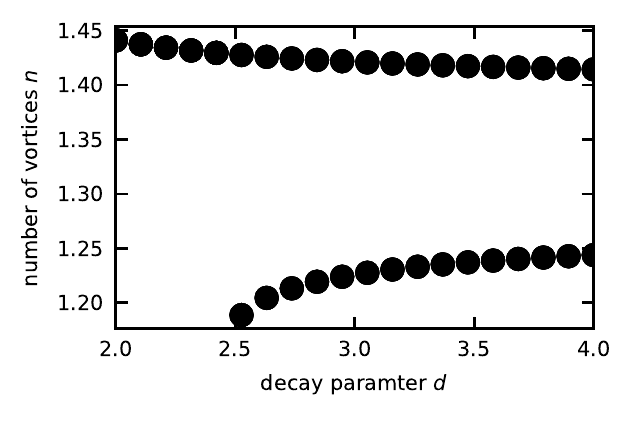}
    \caption{The appearance of the bistable behavior controlled by $d$ (other parameters remain unchanged). The parameter $d$ is related to the cross-section for reconnection of colliding vortices. With decreasing temperature, the vortex motion is less damped thus making the vortices more curved. ``Wiggly" vortices occupy larger area and are therefore more likely to reconnect. The appearance of the bistability with increasing $d$ is therefore in qualitative agreement with the measured temperature dependence.}
    \label{fig:dscaling}
\end{figure}

Finally, the temperature dependence of the experimental observations can be connected, for example, with the parameter $d$, which is related to the cross-section for reconnection of colliding vortices. This will increase with vortex deformation which, in turn, is expected to increase with decreasing temperature \cite{Barenghi1985a}. As shown in Fig.~\ref{fig:dscaling}, the bistability does indeed appear as $d$ increases in qualitative agreement with the data. 

In conclusion, using a microfluidic Helmholtz resonator we have demonstrated a long-lived bistable turbulent behavior in superfluid \4He restricted to quasi-2D channel, which exists below a certain critical temperature. In addition, we observe an unusal backward transition where the flow transitions into a \emph{more} dissipative state as the flow velocity \emph{decreases}. The bistability, hysteresis and the backward transition of the observed turbulence are understood in terms of a model of vortex density as an interplay between spontaneous flow ordering and polarization of turbulence generation. The proposed model is, in principle, applicable to other systems with discrete vorticity (e.g., BECs, superfluid \3He) if the generation of turbulence is in some manner polarized. An interesting question is whether similar behavior is possible in continuous classical systems (indeed, random switching between two degenerate flow configurations has been observed \cite{Woyciekoski2020}). The backward transition is of particular interest, as one usually expects turbulent fluctuations to decrease as the flow driving them is reduced. Considering that the driving mechanisms of, for example, atmospheric or oceanic flows---which are approximately 2D on large scales \cite{Pouquet2013}---are typically not homogeneous and isotropic, the bistable behavior could have implications for weather prediction, climate modelling, and atmospheres of gas giants \cite{Young2017}.

This work was supported by the University of Alberta, Faculty of Science; the Natural Sciences and Engineering Research Council, Canada (Grants Nos. RGPIN-04523-16, DAS-492947- 16, and CREATE-495446-17); and the Canada Foundation for Innovation. We are grateful to G.G. Popowich for technical assistance and F. Souris for the velocity calibration theory.


\newpage
\onecolumngrid
\appendix
\setcounter{figure}{0}
\renewcommand\thefigure{SM\arabic{figure}}

\section*{Supplementary Information}

\section*{Experimental setup and superfluid velocity calculation.}
\label{sec:A-setup}

Flow in the Helmholtz resonators is driven and sensed capacitively using two aluminum electrodes deposited on the top and bottom wall of the device, forming a parallel plate capacitor (see \cite*{Souris2017} for details on the fabrication process). An alternating voltage of amplitude $U_0$ applied to the electrodes of the device causes a periodic deformation of the walls of the basin due to electrostatic force, which pushes the superfluid in and out of the basin through the two side channels and into the bath, thus driving the Helmholtz mode. The Helmholtz resonance is observed as a periodic variation of the capacitance of the device.

\begin{figure}[h!]
  \centering
  \includegraphics{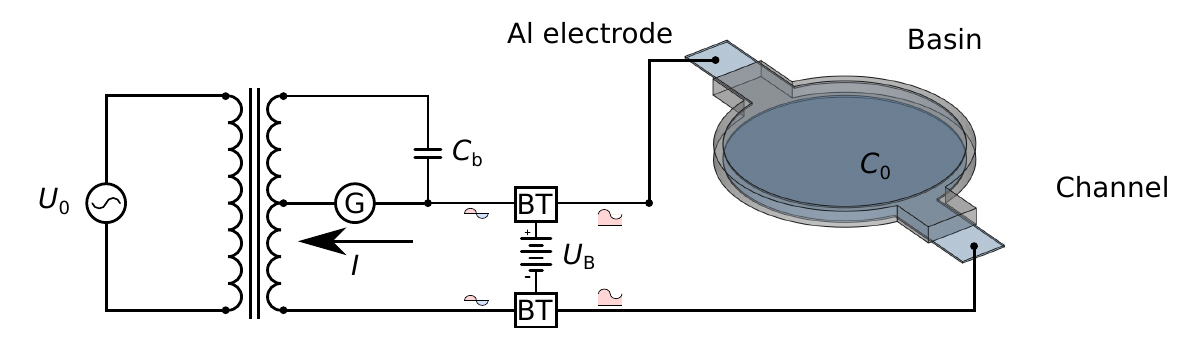}
  \caption{Measurement scheme. Aluminum electrodes of the Helmholtz resonator form a parallel plate capacitor of capacitance $C_0$ biased by the battery $U_\mathrm{B}$. The resonator is wired as one arm of a capacitance bridge (the other arm being the capacitor $C_\mathrm{b}$) which is balanced such that current $I$ through the detector G is approximately zero when flow of the helium is negligible. When the Helmholtz mechanical mode is excited by the oscillating voltage $U_0$, the oscillating pressure in the basin changes the capacitance of the device and thus a nonzero current through the detector G. The bridge circuit is isolated from the battery voltage $U_B$ by the two bias tees (BT).  The transformer ratio is 1:1 on the resonator arm and adjustable for the $C_b$ arm.}
  \label{fig:wiring}
\end{figure}

\begin{figure}
    \centering
    \includegraphics{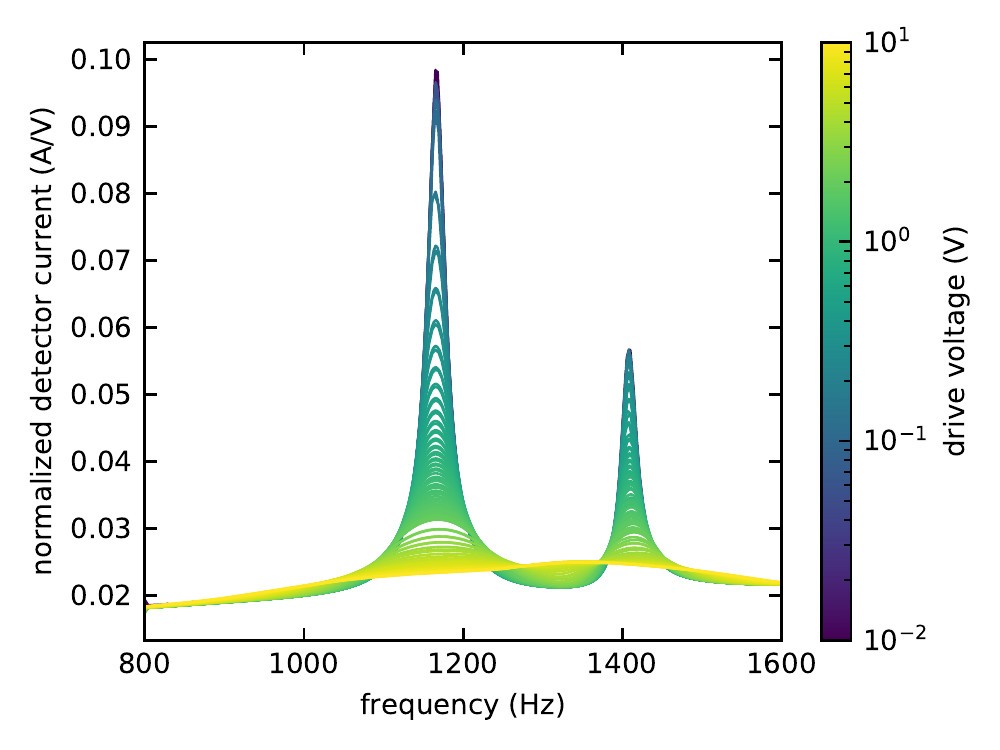}
    \caption{The spectrum of two Helmholtz resonators wired in parallel measured at 1.4 K and normalized by the drive voltage amplitude $U_0$. The fact that the normalized peaks do not overlap for increasing drives indicates nonlinear dissipation. The higher-frequency peak corresponds to the device with $D=1067$ nm and the lower-frequency one to $D = 805$ nm (see Eq.~\ref{eq:omega0}).}
    \label{fig:spectrum}
\end{figure}

The resonator is wired in a bridge circuit shown in Fig.~\ref{fig:wiring}, balanced to the capacitance $C_0$ of the resonator at rest. Change in this capacitance caused by the Helmholtz resonance results in bridge imbalance and a current $I$ through the detector G. An example of the resulting spectrum of two resonators wired in parallel is shown in Fig.~\ref{fig:spectrum}.

The current is first amplified by a transimpedance amplifier (Stanford Research SR570) and then measured by a (Zurich Instruments HF2LI) lock-in amplifier referenced to the frequency of excitation $U_0$. A standard 9V battery is used as the source of the bias voltage ($U_\mathrm{B} = 9.2$ V). The capacitance bridge is the model General Radio 1615-A.

In this section we first analyze the equation of motion of the fluid in channels, approximated as a mass on a spring, and derive the relationship between the oscillating driving voltage $U_0$ and the pressure gradient in the channels. Next, we calculate the superfluid velocity in the channels from the current $I$ through the detector.

\textbf{Helmholtz equation of motion. --} We derive the equation of motion of the superfluid in the Helmholtz resonator with explicit drive and damping forces. We approximate the flow in the channel as a mass on a spring, which is displaced by distance $y$ (positive in the direction away from the basin). The average displacement of the plates of the basin is denoted by $x$ (positive when the basin contracts). We begin by calculating the change in total density of the fluid inside the basin as a response to the mean deformation of the basin $x$ and the displacement of the superfluid inside the channel $y$:
\begin{equation}
  \label{eq:delta-rho}
  \delta \rho = \delta\left(\frac{M_B}{V_B}\right) = \frac{\delta M_B}{V_B} - \frac{M_B}{V_B^2}\delta V_B = \frac{1}{V_B}\left( - 2a\rho_s y + 2\rho A x\right).
\end{equation}
Here $M_B = \rho V_B$ is the mass of the fluid inside the basin, $V_B = AD$ is the volume of the basin ($A$ being its area and $D$ the confinement) and from the assumption that only the superfluid moves $\delta M_B = -2a\rho_s y$ ($a = wD$ is the cross-sectional area of the channel; the factor of 2 comes from the two channels), and $\delta V_B = -2Ax$ is the change in basin volume due to motion of the plates. A change in density corresponds to a change in pressure via the compressibility $\chi$, $\delta \rho = \rho\chi\delta P$, or
\begin{equation}
  \label{eq:delta-P}
  \delta P = \frac{1}{\rho\chi V_B}\left(2\rho A x - 2a \rho_s y\right).
\end{equation}

Balancing forces on the plate (neglecting its inertia) yields
\begin{equation}
  \label{eq:plate-balance}
  \Fes = \frac{1}{2}k_p x + A\delta P,
\end{equation}
where $\Fes = C_0U^2/(2D)$ is the electrostatic force between the parallel plates of the capacitor formed by the circular electrodes in the basin, $U=U_B + U_0$ is the total applied voltage and $k_p = 2.4\times10^7$ N/m \cite*{Souris2017} is the stiffness of the substrate deflection (note that this is double that in Ref.~\cite*{Souris2017}, where the stiffness refers to deflection of both plates in parallel). Expressing $x$ from Eq.~\ref{eq:plate-balance} and substituting back in to Eq.~\ref{eq:delta-P} yields
\begin{equation}
  \label{eq:delta-P-F}
  \delta P = \frac{k_p}{\rho(\chi V_Bk_p + 4A^2)}\left[\frac{4\rho A}{k_p}\Fes - 2a\rho_s y\right].
\end{equation}

The superfluid inside the channel is accelerated by the pressure
\begin{align}
  \label{eq:y-motion}
  \rho_s a l \ddot y &= \frac{\rho_s}{\rho}a\delta P - F_f,\\
  \rho_s a l \ddot y &= \frac{\rho_sk_pa}{\rho^2(\chi V_Bk_p + 4A^2)}\left[\frac{4\rho A}{k_p}\Fes - 2a\rho_s y\right] - F_f,
\end{align}
where we included a friction force $F_f = al\rho_s\zeta\dot y$. Here $\zeta$ is a friction parameter with units of frequency but will remain otherwise unspecified for now. Rearranging,
\begin{equation}
  \label{eq:y-lho}
  \ddot y + \frac{2\rho_s k_p a}{l\rho^2(\chi V_B k_p + 4A^2)} y + \zeta\dot y = \frac{1}{\rho}\frac{4A}{l(\chi V_B k_p + 4A^2)}\Fes,
\end{equation}
from which the resonance frequency follows
\begin{equation}
  \label{eq:omega0}
  \omega_0^2 = \frac{2a}{l\rho}\frac{\rho_s}{\rho}\frac{k_p}{4A^2(1 + \Sigma)},
\end{equation}
where $\Sigma = \chi D k_p/(4A)$.

Finally, the driving pressure gradient, the quantity shown on the x-axis in Fig.~2A,B of the main text, is given by
\begin{equation}
    \frac{\delta P}{l} = \frac{4A}{(\chi V_B k_p + 4A^2)l}\Fes = \frac{4AC_0U_B}{(\chi V_B k_p + 4A^2)Dl}U_0,
\end{equation}
where we take only the component of the force $F_\mathrm{es}$ on resonance with the Helmholtz mode ($U_0$ being the AC drive), $F_\mathrm{es}^\mathrm{res} = C_0U_0U_B/D$.

\textbf{Calculation of velocity from detector current. --} Whenever the capacitance bridge in the measurement circuit shown in Fig.~\ref{fig:wiring} becomes imbalanced, current will flow through the detector. Assuming that the bridge is tuned to the total capacitance of the devices at rest, the current through the detector at the frequency of the drive (the only component detected by the lock-in amplifier) is given only by the oscillation of the device capacitance due to the Helmholtz resonance,
\begin{equation}
  \label{eq:I-C}
  I = \diff{(CU)}{t} = U_B\diff{C}{t} = U_B\diff{C}{y}\dot y,
\end{equation}
where $U_B$ is the bias voltage. The change of capacitance with superfluid displacement in the channel can be written as
\begin{equation}
  \label{eq:dCdy}
  \diff{C}{y} = \frac{C_0}{\varepsilon}\diff{\varepsilon}{\rho}\diff{\rho}{y} + 2\frac{C_0}{D}\diff{x}{y},
\end{equation}
where $C_0 = \varepsilon_0\varepsilon A_\mathrm{el} / D$ is the capacitance of an undisturbed device with $\varepsilon_0,\varepsilon$ being the vacuum permittivity and dielectric constant of helium, respectively, and $A_\mathrm{el}$ the area of the electrodes. The spacing between electrodes is given by $h = D - 2x$, hence the second term in Eq.~\ref{eq:dCdy}.

Neglecting the dependence of polarizability of helium on density \cite*{Kierstead1976} and using the Clausius-Mossoti relation we can estimate the change in dielectric constant as
\begin{equation}
  \label{eq:eps-rho}
  \diff{\varepsilon}{\rho} = \frac{\varepsilon - 1}{\rho}.
\end{equation}
The change of density with superfluid displacement can be calculated directly from Eq.~\ref{eq:delta-P-F} and using $\dd \rho = \rho\chi\dd P$,
\begin{equation}
  \label{eq:rho-y}
  \diff{\rho}{y} = -\frac{2a\rho_s}{V_B}\frac{2\Sigma}{1 + 2\Sigma}.
\end{equation}
where $\Sigma = \chi D k_p / (4A)$. Differentiating Eq.~\ref{eq:delta-rho} with respect to $y$ and putting the result equal to Eq.~\ref{eq:rho-y} results in an equation for $\dd x / \dd y$ and yields
\begin{equation}
  \label{eq:x-y}
  \diff{x}{y} = \frac{a\rho_s}{A\rho}\left(1 - \frac{2\Sigma}{1 + 2\Sigma}\right).
\end{equation}
Inserting Eq.~\ref{eq:x-y}, Eq.~\ref{eq:rho-y} and Eq.~\ref{eq:eps-rho} back into Eq.~\ref{eq:dCdy} yields
\begin{equation}
  \label{eq:dCdy-final}
  g \equiv \frac{1}{C_0}\diff{C}{y} = \frac{2a\rho_s}{V_B\rho}\left(1 - 2\frac{\varepsilon - 1}{\varepsilon}\Sigma\right)\frac{1}{1 + 2\Sigma}.
\end{equation}
Finally, the flow velocity is calculated as
\begin{equation}
  \label{eq:I-V}
  \dot y = \frac{I}{gV_BC_0},
\end{equation}
assuming that the background has been subtracted from $I$.

\subsection{Two-dimensionality of turbulence}
\label{sec:two-dim}

To what extent can the studied flow be considered 2D? The thickness of the flow channel is too large (i.e., $D$ is much larger than the coherence length, $\approx$\AA) for finite-size effects of 2D superfluidity to be relevant \cite*{Bishop1978}. 2D turbulence, on the other hand, requires only that the fluctuating velocity is restricted to 2D. This is essentially controlled by the channel aspect ratio and damping of the self-induced vortex motion.

Turbulence in He-II, especially when forced by a pressure gradient in large systems, typically behaves quasi-classically---as a classical liquid with effective viscosity \cite{Babuin2014}. Thus we first verify that the turbulence could be considered two-dimensional based on classical fluid dynamics criteria.

The forced superflow induced by the Helmholtz resonance is naturally 2D, however, the device geometry (specifically, the sharp corners near where the basin connects to the channel) induces shear on the scale of the channel width $w = 1.6$ mm, which can, in principle, drive a 3D flow instability. It was shown by Benavides and Alexakis \cite*{Benavides2017}, for systems of reduced dimensionality that the direction of the turbulent energy cascade critically depends on the ratio $w/D$ of forcing to confinement scale. Specifically, for $w/D \gtrsim \sqrt{\mathrm{Re}}$ the turbulence develops the 2D inverse energy cascade. Here, $\mathrm{Re}$ is the Reynolds number, which we define for our system using the effective quasi-classical viscosity He-II \cite{Babuin2014} $\nu_\mathrm{eff} \approx 0.1\kappa$ as $\mathrm{Re} = wv_s/\nu_\mathrm{eff}$. In our experiments $w/D \approx 1500$ for $D = 1067$ nm and the highest experimentally achieved $\sqrt{\mathrm{Re}} \approx 400$. Therefore, from a standpoint of classical turbulence, the turbulence in our devices ought to be in the 2D regime. It should be noted, however, that even if a few vertical modes of motion are possible, the inverse energy cascade responsible for appearance of large-scale features is still expected to be present \cite*{Pouquet2017}.

The turbulent fluctuations, however, will be also strongly affected by the presence of quantized vortices, whose core size is on the scale of $a_0 \approx 0.1$ nm---significantly smaller than the confinement imposed by the device. A potential complication arises from vortex pinning on rough surfaces. The RMS surface roughness of our devices is expected to be \cite{Duh2012} about 1 nm, which puts the flow velocity required to dislodge a vortex from a typical surface defect \cite{Schwarz1985} at about 4~cm/s. The velocities we observe in the turbulent regime are significantly higher, thus it is unlikely that pinning plays an important role for our results.

\begin{figure}
    \centering
    \includegraphics{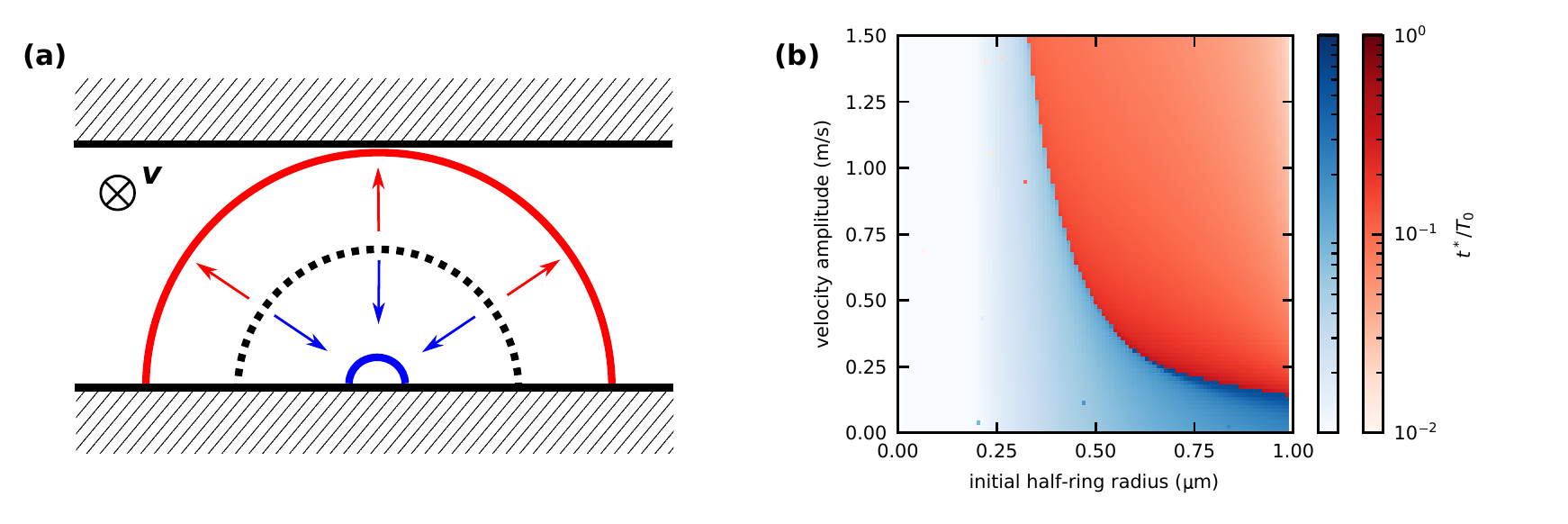}
    \caption{(a) Configuration of the flow and vortex for calculation of the lifetime of the loop. Under the imposed oscillatory flow, the loop can either annihilate or expand, reconnect with the opposite wall and thus create a dipole pair of vortices spanning the confinement. (b) Time, $t^*$ relative to the flow oscillation period $T_0$, for a half ring to grow to the size of the channel (and thus reconnect with opposite wall to form a vortex dipole, red scale) or to completely annihilate (blue scale) in oscillating superfluid flow of varying amplitudes. Flow frequency was assumed to be $f=1200$ Hz ($T_0 = 1/f = 0.83$ ms) and mutual friction constant $\alpha = 0.034$ (corresponding to approximately 1.3 K). Higher $\alpha$ (temperature) results in typically shorter lifetime.}
    \label{fig:loop-lifetime}
\end{figure}

It is in principle possible that a portion of the vortices in the flow are intrinsically three-dimensional, e.g. half-loops pinned on one of the opposing confining walls. To estimate the importance of such vortices we estimate their lifetime in a configuration shown in Fig.~\ref{fig:loop-lifetime}(a). We assume a circular vortex attached to one wall aligned perpendicular to the applied oscillating flow. The self-induced velocity of the ring (neglecting pinning) as a function of its radius is given by \cite{Donnelly1991}
\begin{equation}
    v_i(R) = \frac{\kappa}{4\pi R} \left[\log\left(\frac{8R}{a_0}\right) - \frac{1}{4}\right],
\end{equation}
and, for stationary normal fluid, the change in radius is given by \cite{Donnelly1991}
\begin{equation}
    \dot R = \alpha\left[V_\mathrm{s}(t) - v_i(R)\right],
\end{equation}
where $\alpha$ is the mutual friction constant \cite{Donnelly1998} and $V_\mathrm{s}(t) = V_\mathrm{s0}\sin\Omega t$ is the imposed superflow. We numerically integrate the evolution of $R$ for a range of initial radii $R_0 = R(t=0)$ and velocity amplitudes, terminating the calculation when either $R\approx 0$ and the loop is annihilated or when $R\approx D = 1$ $\mathrm{\mu}$m and the loop reconnects with the opposing wall, thus transforming into a vortex dipole. As shown in Fig.~\ref{fig:loop-lifetime}(b), the typical lifetime $t^*$ of half-rings for the parameters typical of our experiment is shorter than the flow oscillation period $T_0$, reaching, at most, about 0.6$T_0$ for a very specific choice of parameters. Vortex loops attached to a surface are thus short-lived transient objects. Creation and expansion of these loops is a likely scenario for vortex splitting and unpolarized injection, which feature in the quasi-2D model of Eqs.~(3,4) of the main text, discussed further in the next section.

Note, however, that we neglected the effects of the opposing wall on the self-induced velocity of the ring. This will cause the vortex to deform and be attracted to the opposing wall, thus slightly altering the lifetime. Changing the phase of the oscillating flow either does not significantly influence the outcome or causes loops of all sizes to quickly decay. Changing the angle between the plane of the loop and flow velocity would result in a somewhat more complicated transient flow, which is, however, unlikely to terminate in a significantly different manner.

The calculation outlined above is not valid for vortex loop radii $R$ comparable with the surface roughness $b\approx 1$ nm, since the vortex will be subject to highly nonuniform flow resulting from the surface imperfections. Stagg et \emph{al.} \cite{Stagg2017} studied a flow close to an irregular surface using a simulation of vortices in a Bose-Einstein condensate in the zero-temperature limit using the Gross-Pitaevskii equation (GPE). In that work, a dense layer of vortices was found in the rough landscape of the surface sustained by intrinsic nucleation of vortices on the protruding peaks of the surface. It is possible that such a dense boundary layer exists in our case as well, however, results obtained using GPE ought to be adopted with caution for helium at finite temperatures. Intrinsic nucleation of vortices in He-II requires significantly higher velocities and mutual friction at finite temperatures will strongly damp any small, highly-curved vortex structures. Regardless, this boundary layer is expected to be confined to within the scale of surface roughness \cite{Stagg2017} which in our case is about 0.1\% of the confinement thus making it unlikely to be a significant contribution to the observed macroscopic drag.

Finally, the vortices connecting the two confining walls can, in principle, deform arbitrarily on the scale of the confinement, $D\approx 10^4 a_0$. We estimate the dynamical importance of these deformations by comparing their typical rate of decay to the time scale of their forcing, i.e., the flow oscillation period. Assume that the vertical modes of flow, mediated by the vortex deformation, take the form of a cascade of Kelvin waves---helical wave modes on vortices \cite*{Donnelly1991}. The decay rate of a Kelvin wave mode of wave vector $k$ is $\tau = {\kappa}/{4\pi\alpha k^2}$ \cite*{Barenghi1985a}. The smallest admissible $k \approx 2\pi/D$ results for $T = 1.3$ K and $D=1067$ nm in $\tau \approx 30$ $\mu$s.  Increasing temperature will decrease $\tau$. The decay rate of Kelvin waves is thus significantly faster than the time scale of their pumping (i.e., flow period, which is of the order of 1 ms) and comparable to the inverse frequency of the Kelvin mode itself \cite*{Donnelly1991}, i.e., no Kelvin wave cascade is likely to develop along the individual vortices since the largest scales are already in the dissipative range. Other modes of vortex deformation (e.g.,  solitons \cite*{Hopfinger1982}), which cannot be decomposed to Kelvin waves, are possible. However, since the local velocity of the deformed line, and thus its decay rate mediated by mutual friction, are primarily determined by the local curvature, we expect the decay of these deformations to be comparable to that of the Kelvin waves. The amplitude of thermally excited Kelvin waves is also expected to be negligible \cite{Barenghi1985a}. We therefore consider the vortices in our system which span the confinement to be point-like. Decreasing the temperature, particularly below 1 K, would suppress the mutual friction damping and allow the vortices to deform strongly. Therefore we expect the turbulence to cease to be 2D-like at sufficiently low temperatures.

\subsection{Vortex density model}
\label{sec:model}

Discrete quantized vortices are transported by flow similar to how vorticity is transported in classical 2D flow \cite*{Kraichnan1980}:
\begin{equation}
    \frac{\partial n_\pm}{\partial t} + (\mathbf{v}_\mathrm{s}\cdot\nabla)n_\pm = (n_+ + n_-)b(\mathbf{v}_\mathrm{s}) - dn_+n_- + g_\pm,
\end{equation}
where the terms on the right hand side correspond, respectively, to splitting, decay by collision and generation of vortices by the external drive. Note that this is not simply a passive scalar transport with source terms, since $\mathbf{v}_\mathrm{s}$ depends on the vortex distribution. We expect the splitting rate $b$ to depend on velocity, possibly exhibiting critical behavior itself. For simplicity, however, we take $b$ to be constant since in a high-velocity regime it is likely to be dominated by the flow oscillation period, which is independent of velocity.

Averaged over the flow oscillation period, the advection term $(\mathbf{v}_\mathrm{s}\cdot\nabla)n_\pm$ will have no effect on the vortex density far from the system boundaries. In the region near the boundaries, however, some vortices will be transported toward the wall and annihilated, i.e., the average effect of the advection is to reduce the vortex number. Since the vortex density will vary on the scale of the channel width, we approximate the gradient term as $(\mathbf{v}_\mathrm{s}\cdot\nabla)n_\pm \approx v_\mathrm{s}n_\pm/w$. Putting $a = b - pv_\mathrm{s}/w$, where $p$ characterises the inhomogeneous distribution of vortices throughout the channel, we recover Eqs.~1,2 of the main text. The assumption of velocity-independent $b$ and $d$ limits the applicability of the model to turbulent states at relatively high velocity and makes it unsuitable for modelling transition to turbulence from the laminar state or predicting the scaling of vortex number with velocity.

Following from Eqs.~1, 2 of the main text, the total vortex density $n = n_+ + n_-$ and polarization $s = (n_+ - n_-)/n$ obey
\begin{equation}
  \label{eq:dndt-S}
  \diff{n}{t} = (a+b)n - \frac{1}{2}dn^2(1 - s^2) + g,
\end{equation}
and
\begin{equation}
  \label{eq:dsdt-S}
  \diff{s}{t} = -2bs + \frac{1}{2}dns(1 - s^2) + \frac{g_s}{n},
\end{equation}
where we grouped all terms depending on $g_\pm$ to new terms $g = g_+ + g_-$ and $g_s = (1-s)g_+ - (1+s)g_-$. The generation terms $g_\pm$ in Eqs.~1, 2 of the main text represent extrinsic or intrinsic nucleation of vortices and are likely to be concentrated near the sharp corners connecting the basin and the channel. The vortices generated at these edges in a polarized configuration are advected into the channel where they contribute to the observed drag. Near the corners, however, the polarization $s$ will likely be dominated by the instantaneous flow and, averaged over the flow oscillation period, $s\approx 0$ making the $g_s \approx g_+ - g_-$, independent of $s$. For simplicity we adopt $g$ and $g_s$ as independent control parameters, rather than $g_\pm$. It should be noted, however, that it is the assumption of $s$-independent $g_s$ that allows for bistable solutions.

The equations above are assumed to be local, but spatially averaged quantities are required for comparison with the experiment. The total vortex density $n$ is an always positive quantity and thus, to a first approximation, can be replaced by its spatial average. The vortex polarization $s$, on the other hand, has a vanishing average since we assume that the flow will remain on average neutral.

\begin{figure}
    \centering
    \includegraphics{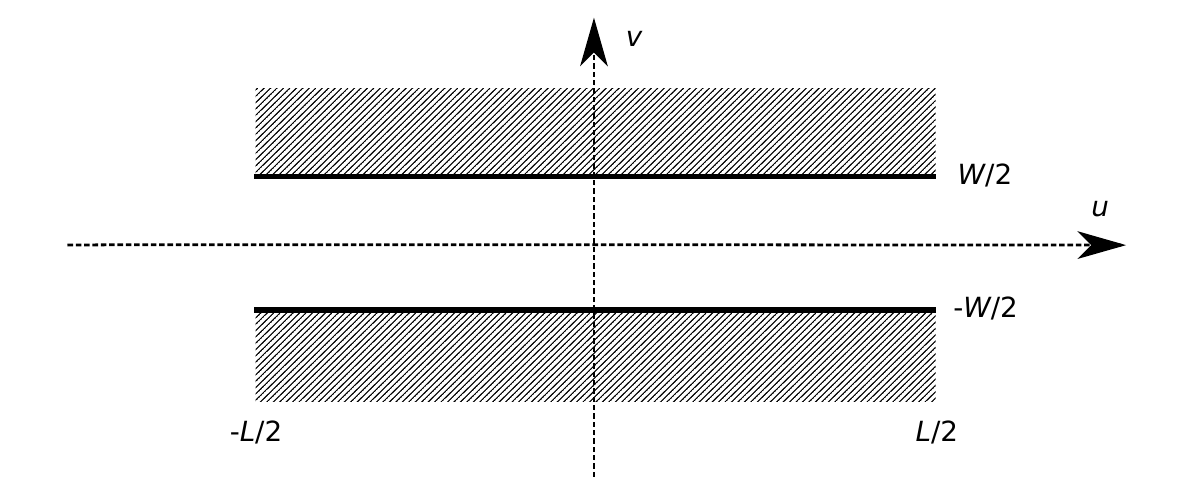}
    \caption{Simplified geometry for the decomposition of $s$ into orthogonal modes.}
    \label{fig:simple-geometry}
\end{figure}

To connect Eq.~\ref{eq:dsdt-S} to averaged quantities, let us consider the simplified device geometry shown in Fig.~\ref{fig:simple-geometry}. The basin is removed and a single channel runs through the entire length of the device, but otherwise we assume general flow features similar to the real device (e.g., flow direction, behavior of the generation terms $g,g_s$). From the symmetry of the problem, $s$ is anti-symmetric with respect to mirroring about either of the axes and thus can be decomposed into orthogonal modes as
\begin{equation}
    \label{eq:s-expansion}
    s(u, v) = \sum_{k,l} s_{kl}\sin\left(\frac{2\pi k}{L}u\right)\sin\left(\frac{2\pi l}{W}v\right),
\end{equation}
and the spatial average is then given by $\langle s^2\rangle = 1/4\sum_{kl}|s_{kl}|^2$. In the actual device geometry the modes in the expansion Eq.~\ref{eq:s-expansion} will be more complicated but could, in principle, be constructed by a suitable transformation of the rectangular domain of Fig.~\ref{fig:simple-geometry} onto the actual device geometry. Truncating the expansion at the lowest $s_{11}$ mode (which is likely to be the dominant term in the generation $g_s$) allows us to essentially use Eqs.~\ref{eq:dndt-S}, \ref{eq:dsdt-S} as they are and recover the results from the main text.

\begin{figure}
    \centering
    \includegraphics{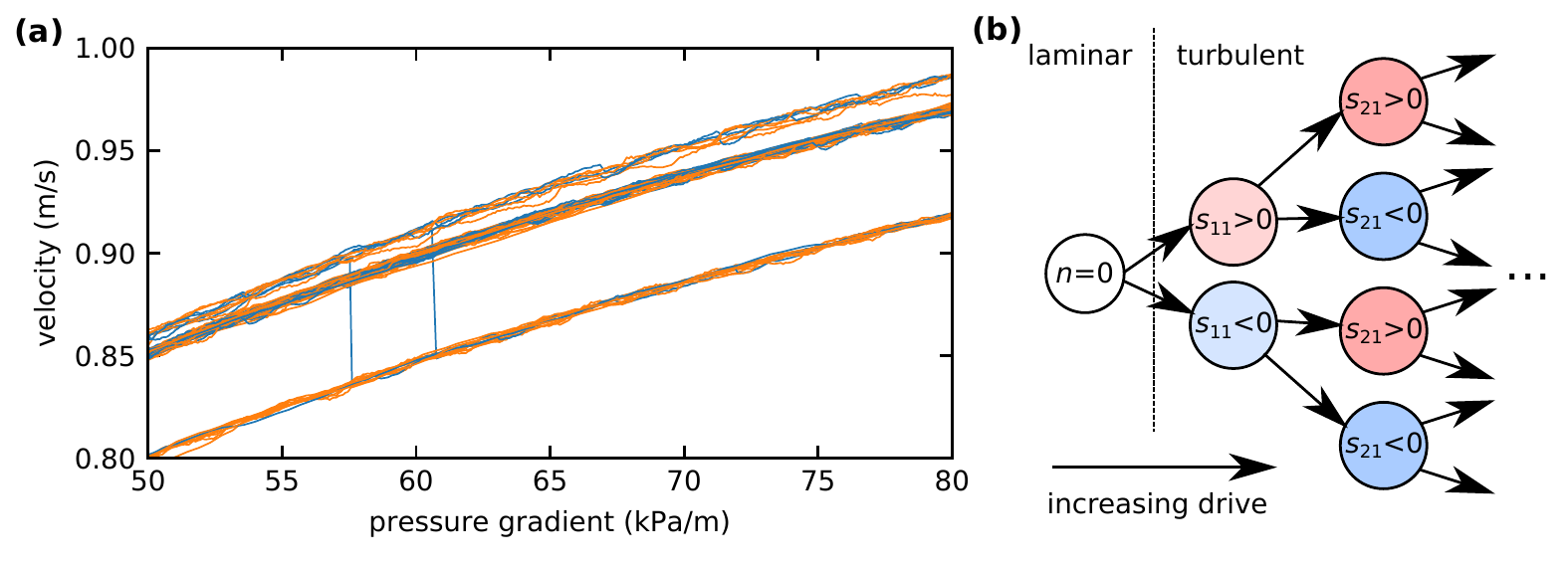}
    \caption{(a) Three distinct turbulent branches of the pressure-velocity curve at 1.4 K (the laminar regime, see Fig.~2a,b of the main text, is not shown in this plot). The multi-stable behaviour hints at the involvement of higher modes of the vortex polarization. (b) An illustration of a tree of multi-stable states generated by the Eqs.~\ref{eq:dndt-S}, \ref{eq:dsdt-S}.}
    \label{fig:multi-stability}
\end{figure}

In principle higher modes $s_{kl}$ can be considered, where Eq.~\ref{eq:dsdt-S} would be replaced by a set of equations for each mode coupled through nonlinear terms. The generation term $g_s$ is unlikely to have a single-mode decomposition and the nonlinear terms (in $s$) in Eq.~\ref{eq:dsdt-S} will excite higher modes at the expense of lower modes. This picture is fully consistent with the forward enstrophy (quadratic integral of vorticity) cascade of classical 2D turbulence \cite*{Kraichnan1980}. Higher modes will again exhibit near-degeneracy of the $s_{kl} > 0$ and $s_{kl} < 0$ solutions lifted by the appropriate mode of the $g_s$ and, possibly, by the lower-lying modes through the nonlinear terms. This will result in more general multi-stability of the mean vortex number $n$, as illustrated in Fig.~\ref{fig:multi-stability}b. As the flow velocity or drive increases, the system will randomly select either the $s>0$ or $s<0$ solution. As the drive increases further, the higher-order terms will become important, which will again be selected randomly, splitting each branch further. The beginning of this tree of turbulent states is, perhaps, already seen in the high-velocity part of the pressure-velocity curves at 1.4 K shown in Fig.~2a of the main text and highlighted in Fig.~\ref{fig:multi-stability} where three distinct turbulent branches are clearly seen.

\subsection{Comparison of turbulence in 805 nm and 1067 nm confinements}
\label{sec:805nm}

\begin{figure}
    \centering
    \includegraphics{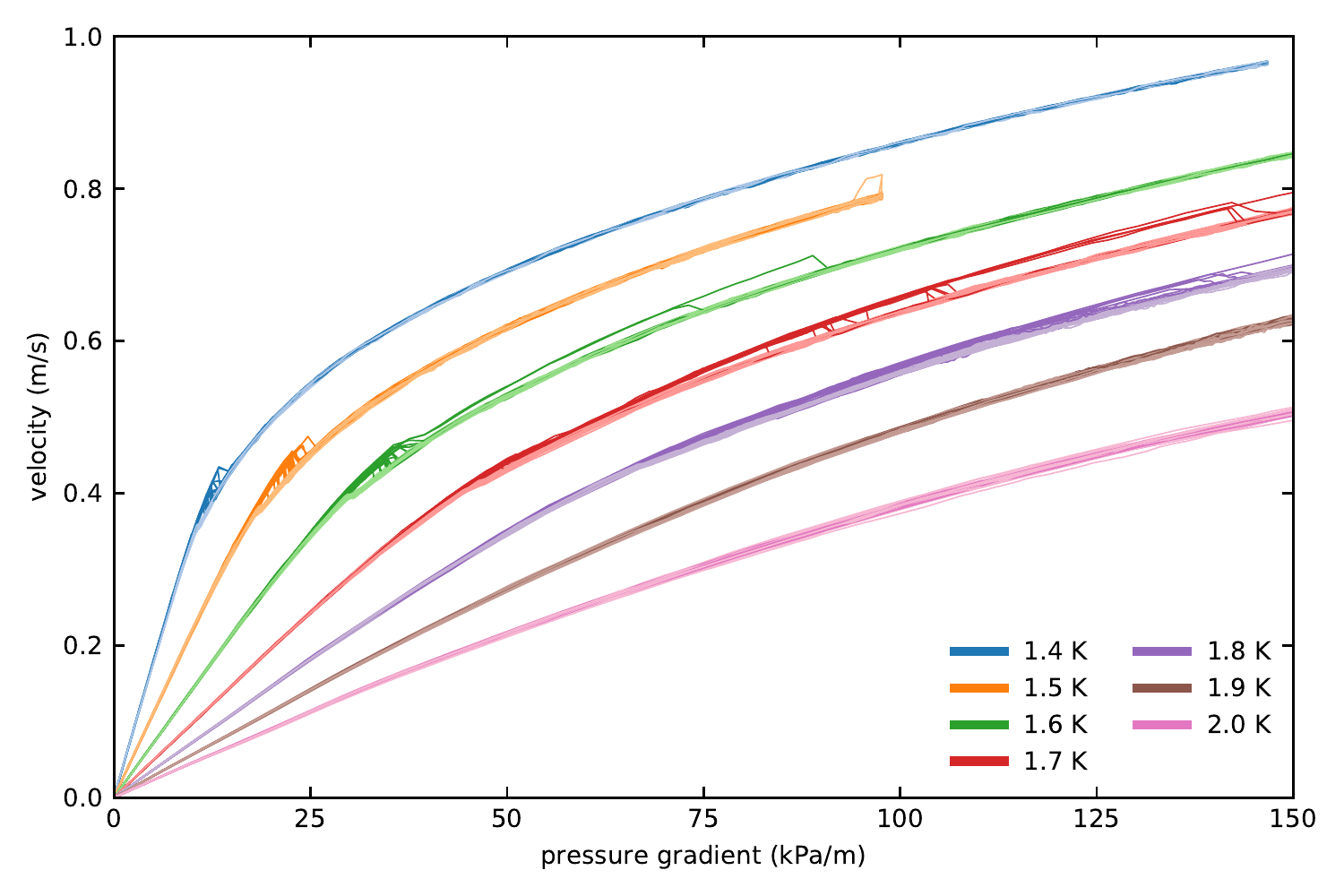}
    \caption{Velocity-pressure gradient relationship for the 805 nm confinement, under identical conditions as 1067 nm confinement shown in Fig.~2 of the main text and Fig.~\ref{fig:fv-1000}. Darker curves show increasing pressure gradient, lighter decreasing. The bistability is observed here in a weaker form between 1.6 and 1.8 K.}
    \label{fig:fv-800}
\end{figure}

\begin{figure}
    \centering
    \includegraphics{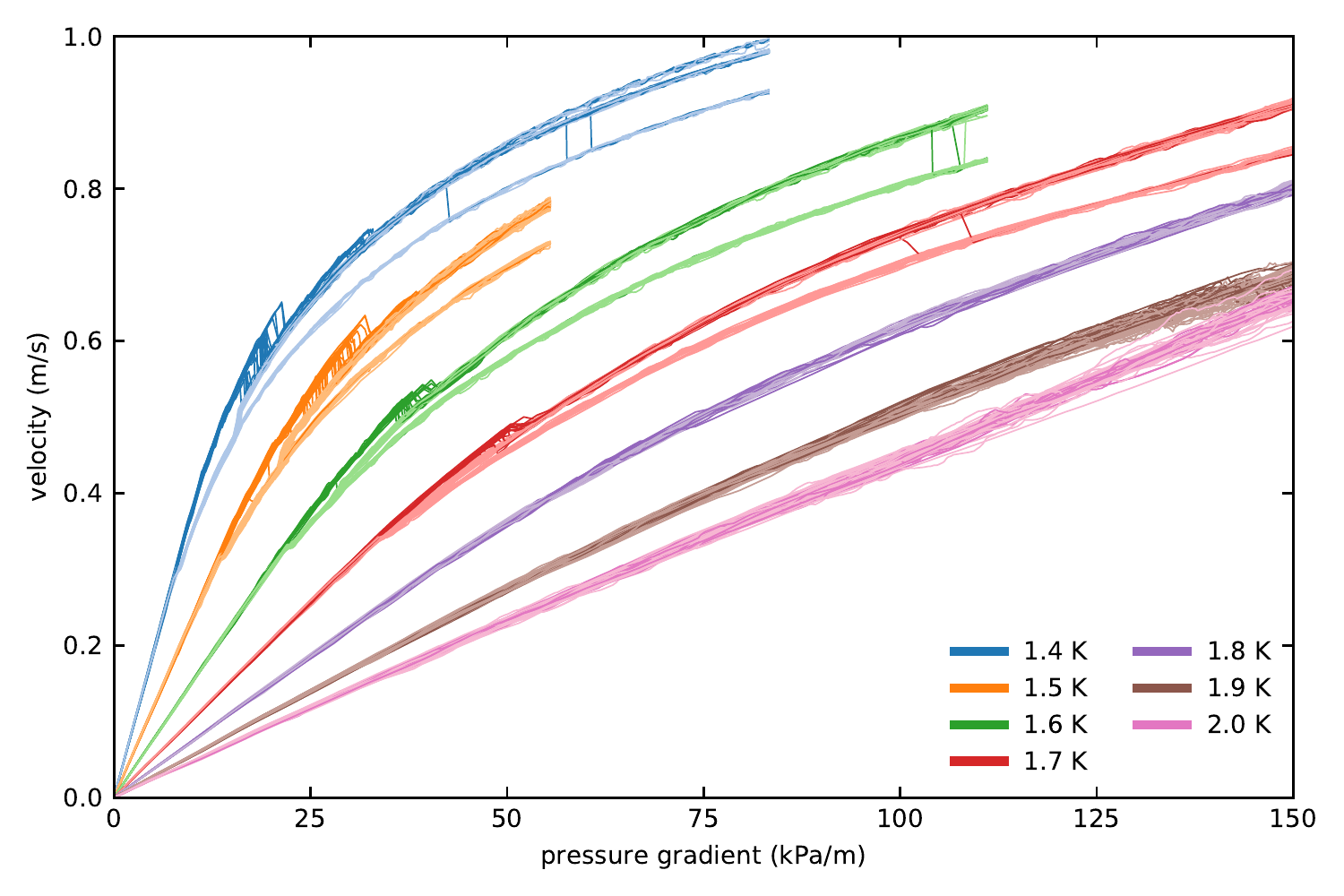}
    \caption{Velocity-pressure gradient relationship for the 1067 nm, same data as in  Fig.~2 of the main text shown for a wider range of the pressure gradient. Darker curves show increasing pressure gradient, lighter decreasing. A few random transitions from the less-dissipative to the more-dissipative state are visible for the bistable turbulence at $T < 1.8$ K. The probability of obtaining the more-dissipative state in the high-velocity regimes can be improved by suddenly increasing the velocity from zero to the target velocity, without the preceding slow ramp-up. These measurements are included within the ramp-down sets of curves. We have never observed a transition from the more-dissipative to the less-dissipative state.}
    \label{fig:fv-1000}
\end{figure}

The velocity-pressure gradient curves for 805 nm confinement (shown in Fig.~\ref{fig:fv-800}) were measured in parallel with the 1067 nm confinement (shown in Fig.~\ref{fig:fv-1000} and Fig.~2 of the main text) under identical conditions. The bistability is again present in the 805 nm confinement, but in a weaker form and in the temperature range of 1.6--1.8 K. The bistability also tends to be suppressed at high velocities.

Within the model of Sec.~\ref{sec:model} the bistability can be destroyed in several ways, apart from the already discussed temperature dependence of the decay parameter $d$. For example, increasing the splitting rate $b$ above a critical value will result in a single solution with $s\approx 0$ (i.e., frequent splitting will completely mix the flow). Similarly, increasing $g_s$ above a critical value will destabilise the solution with sign of $s$ opposite to the sign of $g_s$ (i.e., opposing polarization will be overwhelmed by the strong drive). Additionally, the confinement of the device likely affects the $d$ parameter as well -- the smaller 805 nm device allows for lesser lateral deformation of the vortices and hence lowers the effective cross-section for collision, thus reducing $d$, which tends to reduce the bistability.

In fact the bistability is not necessarily destroyed completely. If the environmental disturbances (e.g., vibration of the cryostat) are non-negligible compared to the relative stability of the less-stable state (controlled, for example, by the $d$ parameter), the flow would stochastically transition to the more stable state whenever a sufficiently strong fluctuation randomly occurs. This scenario is consistent with the fact that the transition between the two turbulent states in the temperature range 1.6--1.8 K for the 805 nm confinement does not appear to have a well defined critical velocity.

The device-dependence of the $d$ parameter discussed above, however, does not account for the complete lack of bistability at lower temperatures in the 805 nm device. One possibility for this observation is that critical velocity of type II (in Fig.~2(b) of the main text) moved beyond the critical velocity of type I. Once the laminar flow becomes unstable, only one turbulent state would be available which would thus be the only state observed. Indeed, this would be consistent with a relatively narrow hysteretic region at low temperatures in Fig.~\ref{fig:fv-800}. Additionally, Fig.~2(c) of the main text could suggest that the closing of the gap between critical velocities of types I and II is plausible even for the 1067 nm device at lower temperatures. However, due to lack of data from lower temperatures and lack of a model of the critical velocities this scenario ought to be regarded as a speculation at this point.

In order to describe the destruction of bistability precisely, significantly more detailed understanding of the critical velocities and the vortex-boundary interaction, and the parameters of the Eqs.~\ref{eq:dndt-S}, \ref{eq:dsdt-S} that stem from it, would be required. This will depend, for example, strongly on the morphology of the surface \cite*{Stagg2017} and is beyond the scope of this work.

\end{document}